\documentclass[runningheads]{llncs}
\usepackage{graphicx}
\usepackage{bm}
\usepackage{amsmath}
\usepackage{amsfonts, amssymb}
\usepackage{multirow}
\usepackage{booktabs}
\usepackage{subfigure}
\usepackage{marvosym}
\usepackage{setspace}

\begin{document}
\title{FAN: Fatigue-Aware Network for Click-Through Rate Prediction in E-commerce Recommendation}

\titlerunning{FAN}

\author{
    Ming Li\inst{*} \and
    Naiyin Liu\thanks{Both authors contributed equally. Xiaofeng Pan is the corresponding author.} \and
    Xiaofeng Pan(\Letter) \and
    Yang Huang \and
    Ningning Li \and
    Yingmin Su \and
    Chengjun Mao \and
    Bo Cao
}
\authorrunning{M. Li et al.}
%
\institute{
    Alibaba Group, Hangzhou, China \\
    \email{hongming.lm@alibaba-inc.com, liuny5@mail2.sysu.edu.cn, pxfvintage@163.com, 21631140@zju.edu.cn, \{lily.lnn,yingmin.sym,chengjun.mcj,zhizhao.cb\}@alibaba-inc.com}
}

\maketitle              

\begin{abstract}
Since \emph{clicks} usually contain heavy noise, increasing research efforts have been devoted to modeling implicit negative user behaviors ($i.e.$, \emph{non-clicks}). However, they either rely on explicit negative user behaviors ($e.g.$, \emph{dislikes}) or simply treat \emph{non-clicks} as negative feedback, failing to learn negative user interests comprehensively. In such situations, users may experience fatigue because of seeing too many similar recommendations. In this paper, we propose Fatigue-Aware Network (FAN), a novel CTR model that directly perceives user fatigue from \emph{non-clicks}. Specifically, we first apply Fourier Transformation to the time series generated from \emph{non-clicks}, obtaining its frequency spectrum which contains comprehensive information about user fatigue. Then the frequency spectrum is modulated by category information of the target item to model the bias that both the upper bound of fatigue and users' patience is different for different categories. Moreover, a gating network is adopted to model the confidence of user fatigue and an auxiliary task is designed to guide the learning of user fatigue, so we can obtain a well-learned fatigue representation and combine it with user interests for the final CTR prediction. Experimental results on real-world datasets validate the superiority of FAN and online A/B tests also show FAN outperforms representative CTR models significantly.

\keywords{Recommender System \and Click-Through Rate Prediction \and User Fatigue.}
\end{abstract}
\section{Introduction} 
Recommender Systems (RS) are becoming increasingly indispensable to help users discover their preferred items in situations of information overload, therefore improving the user experience and delivering new business value \cite{xu2015customer,zhang2020empowering}. Typically, an industrial e-commerce recommender system consists of matching and ranking. The matching stage aims to retrieve candidate items related to user interests \cite{sarwar2001item,covington2016deep}, after which the ranking stage predicts precise probabilities of users interacting with these candidate items, $e.g.$, Click-Through Rate (CTR) \cite{zhou2018deep} and Conversion Rate (CVR) \cite{wen2020entire,pan2022metacvr}. In this paper, we focus on the Click-Through Rate (CTR) prediction task of the ranking stage.

Most existing CTR methods \cite{guo2017deepfm,zhou2018deep,CoreAi2022} in e-commerce recommendation mainly rely on implicit positive feedback ($i.e.$, \emph{clicks}) as a positive label and infer users' current interests since clicks can be easily collected in practice. However, click behaviors usually contain heavy noise \cite{wang2018modeling,wen2019leveraging} since there are gaps between \emph{clicks} and users' real preferences, and outdated interests may exist in historical user behaviors \cite{li2020deep}. Moreover, positive feedback is biased toward the choices that the RS offers to its users, as clicks can only be done on items that are exposed to users. As a result, simply focusing on implicit positive feedback will lead to biased modeling of user interests and homogeneous and myopic recommendations, which may harm user experiences \cite{zhao2018explicit}.

Recently, several researchers \cite{gong2022positive,lv2022xdm} notice the drawbacks of merely relying on implicit positive feedback and attempt to leverage the more abundant implicit negative feedback ($i.e.$, \emph{non-clicks}) to learn negative user interests. The key idea is to introduce regularization in the loss function to enforce that the representation of positive feedback should be far away from the \emph{non-clicks}. However, a user doesn't click an item does not necessarily mean the user doesn't like the item. Maybe some similar items are displayed nearby, or maybe the exposed items are simply not well noticed. Therefore, ignoring noise in \emph{non-clicks} may lead to conflicts when modeling user interests and result in inaccurate recommendations. Another line of works \cite{xie2021deep,bian2021denoising,wu2022feedrec} tries to make use of explicit negative feedback ($e.g.$, \emph{dislikes}) to distill negative user interests from implicit feedback ($i.e.$, \emph{non-clicks} and \emph{clicks}). However, explicit negative user behaviors are extremely scarce in e-commerce RS. Less than 0.01\% of impressions will result in \emph{dislikes}, which are less than one-tenth of \emph{purchases}, according to statistics of our e-commerce platform.

In such situations, models are incapable of learning negative user interests comprehensively and users may experience fatigue due to seeing too many similar recommendations. One way to handle fatigue is to recall more new items at the matching stage, which usually doesn't take effect directly because most existing CTR methods are not friendly to items that users didn't see before. Another common solution is the explore and exploit paradigm \cite{auer2002using,li2010contextual} that considers multiple factors including relevance, novelty, and fatigue \cite{ma2016user,cao2020fatigue}. However, they simply model user fatigue by statistics such as the number of similar items shown before but ignore the time-frequency distribution of similar recommendations, which causes the loss of information. Besides, it is believable that user interests and user fatigue can affect each other, so the trade-off between exploration and exploitation may not be an optimal choice.

Based on these observations, we propose a novel CTR model: Fatigue-Aware Network (FAN) which can perceive user fatigue from \emph{non-clicks} more comprehensively, therefore achieving improvements in both CTR prediction and user experience. Naturally, we believe that user fatigue should be modeled at the category level because the item level is too fine-grained and the e-commerce RS usually avoids recommending items recommended recently. Given a pair of user and item, we extract \emph{non-clicks} of the same category in recent days and compute the number of this category in each recommendation request, generating a time series that contains rich information about user fatigue. Then we devise a Fatigue Representation Module (FRM) which applies Fast Fourier Transformation (FFT) \cite{soliman1990continuous} to the time series to obtain its frequency spectrum, which contains comprehensive information about time-frequency distribution of similar recommendations. Considering the bias that both the upper bounds of fatigue and users' patience are different for different categories ($e.g.$, digital products and clothes), we propose to modulate the frequency spectrum by category information. Moreover, a gating network is adopted to model the confidence of user fatigue according to user activeness and an auxiliary task is designed to guide the learning of FRM. In this way, we can obtain a well-learned fatigue representation. At last, user fatigue is incorporated with user interests to make the final CTR prediction so we can avoid too many similar recommendations and achieve more accurate predictions. Our main contributions are summarized as follows:
\vspace{-0.2cm}
\begin{itemize}
    \item We investigate the difficulties of modeling implicit negative user behaviors ($i.e.$, \emph{non-clicks}) for e-commerce recommendation, and propose to model user fatigue explicitly in CTR prediction to avoid users from seeing too many similar recommendations.
    \item We propose a novel FAN model that directly perceives user fatigue from \emph{non-clicks}. Benefiting from the frequency-domain representation and category modulation, we are capable of modeling user fatigue comprehensively and accurately. Besides, with an elaborated auxiliary task, FAN can pay attention well to what users are not interested in.
    \item Experiments on real-world datasets demonstrate the superiority of our FAN model over representative methods, and online tests further show that FAN not only improves model performance but also brings better user experiences. We also conduct extensive analyses to confirm the effectiveness of our design for modeling user fatigue. The code is publicly available\footnote{https://github.com/AaronPanXiaoFeng/FAN}.
\end{itemize}
\vspace{-0.6cm}

\section{Proposed Method}

\begin{figure*}[!pt]
\vspace{-1.5em}
    \centering
    \includegraphics[width=0.9\linewidth]{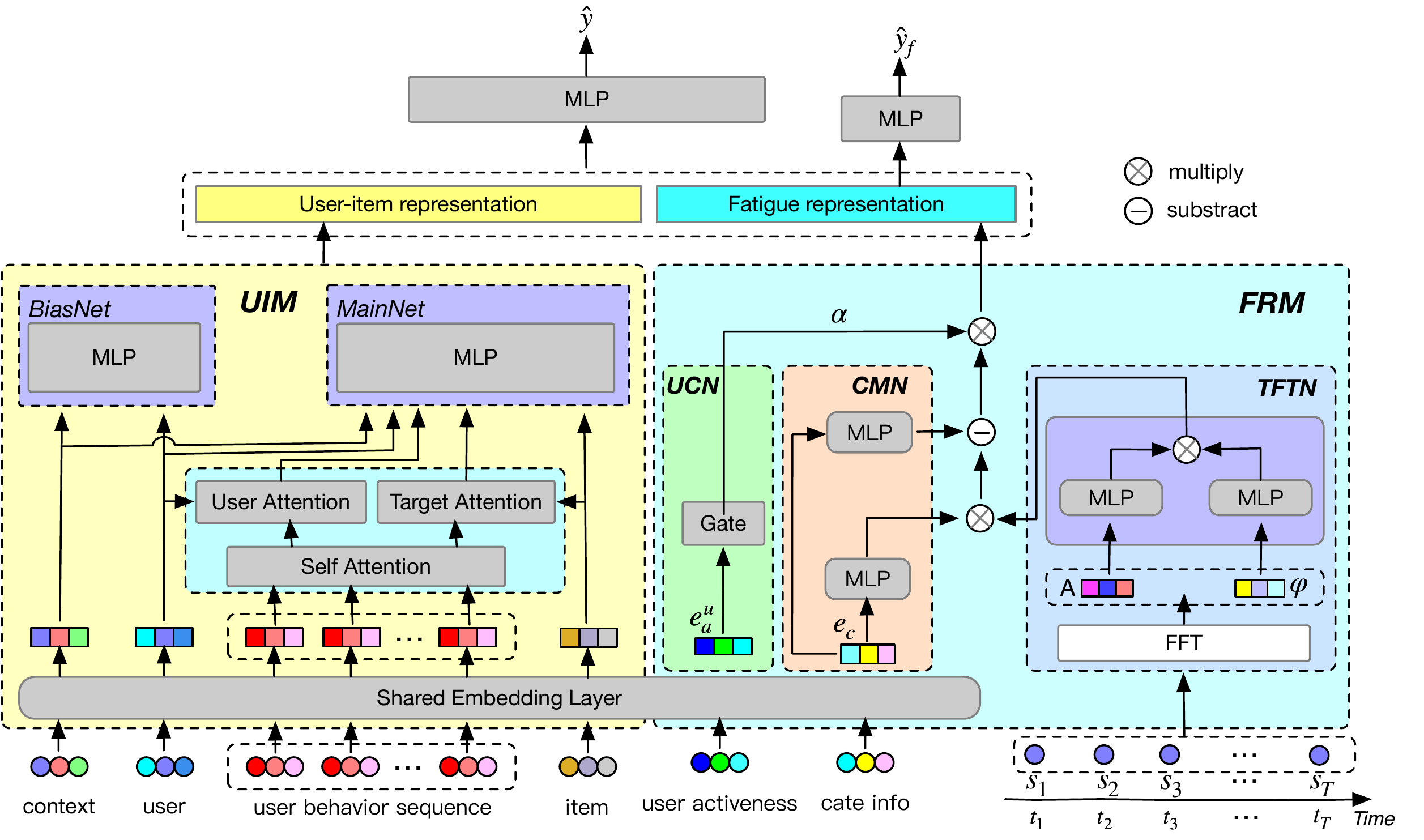}
     \vspace{-1.0em}
    \caption{Framework of the proposed Fatigue-Aware Network (FAN), which consists of the User Interest Module (UIM) and Fatigue Representation Module (FRM). The FRM includes three key components: Time-Frequency Transformation Net (TFTN), Category Modulation Net (CMN), and User Confidence Net (UCN).}
    \label{fig:FAN}
    \vspace{-1.5em}
\end{figure*}

\subsection{Model Input} \label{subsec:input}
In CTR prediction, the model takes input as $(\bm{x}, y) \sim (X,Y)$, where $\bm{x}$ is the feature and $y \in \{0,1\}$ is the click label. Specifically, the features in this work consist of five parts: 1) user behavior sequence $\bm{x}^{seq}$; 2) user features $\bm{x}^u$ including the user profile and user statistic features; 3) item features $\bm{x}^i$ such as item id, category, brand, and related statistic features; 4) context features $\bm{x}^c$ such as position and time information; 5) fatigue time series $\bm{S}$, $i.e.$, a sequence of statistics of \emph{non-clicks} on target category, which is first proposed in this work and will be detailed below.

Now we describe the extraction process of fatigue time series. Given a pair of user and item, we first retrieve recommendation requests of the user in recent days and order them by time. For each request, we compute the number of \emph{non-clicks} which has the same category as the target item, obtaining a time series $\bm{S}=\{s_1, s_2, ..., s_T\}$, where $s_T$ denotes the number of \emph{non-clicks} on the target category in $T$-th request and $T$ is the number of requests. Elements in $\bm{S}$ can be regarded as raw indicators of fatigue in the time domain. To ensure the correlation between $\bm{S}$ and fatigue, we adopt the following rules for special cases in the extraction process: 1) If there are clicks on the target category in the $i$-th request, we set $s_i$ to $0$ because it's more related to the user's positive interest and contributes less to fatigue; 2) If there is no item of the target category in a request, we ignore this request because it is not relevant to the target fatigue. By aggregating \emph{non-clicks} on request granularity and organizing statistics in order of time, the raw information about fatigue is more confident and comprehensive.

\subsection{User Interest Module}
As shown in Figure~\ref{fig:FAN}, the input features of UIM consist of $\bm{x}^u$, $\bm{x}^i$, $\bm{x}^c$ and $\bm{x}^{seq}$, which are detailed in Section~\ref{subsec:input}. They can be further grouped into two kinds of features: categorical features and numerical features. We discrete the numerical features based on their boundary values and transform them into the categorical type. Then each categorical feature is encoded as a one-hot vector. Due to the sparseness nature of one-hot encoding, they are further processed by a Shared Embedding Layer so we obtain the embedded user features, item features, context features, and user behavior sequence, $i.e.$, $\bm{e}^u$, $\bm{e}^i$, $\bm{e}^c$ and $\bm{e}^{seq}=\{\bm{e}^i_1,...,\bm{e}^i_l\}$, where $\bm{e}^i_l$ denotes the item embedding of $l$-th user behavior and $l$ is the sequence length.

For user behavior sequence, we perform three kinds of attention calculation. Firstly a multi-head self-attention \cite{vaswani2017attention} is calculated over $\bm{e}^{seq}$ to model user preference from multiple views of interest and $\bm{\hat{e}}^{seq}=\{\bm{\hat{e}}^i_1,...,\bm{\hat{e}}^i_l\}$ is the output. Secondly, user attention $\bm{a}^u$ is calculated to mine personalized information with $\bm{e}^{u}$ attending to $\bm{\hat{e}}^{seq}$. Thirdly, target attention $\bm{a}^i$ is employed to activate historical interests related to the target item with $\bm{e}^{i}$ attending to $\bm{\hat{e}}^{seq}$. To preserve the original information for further learning of interactions, we concatenate $\bm{e}^{i}$, $\bm{e}^{u}$ and $\bm{e}^{c}$ with $\bm{a}^u$ and $\bm{a}^i$, and feed them into the MainNet, $i.e.$, a Multi-Layer Perception (MLP). Meanwhile, we feed $\bm{e}^{u}$ and $\bm{e}^{c}$ into another MLP ($i.e.$, the BiasNet) to model the bias that different users in different contexts usually behave differently even to similar items. Finally, the outputs of MainNet and BiasNet are concatenated to obtain the user-item representation $\bm{r}_{ui}$.

\subsection{Fatigue Representation Module} \label{FRM}
\subsubsection{Time-Frequency Transformation Net.} \label{TFTN}
Given the fatigue time series $\bm{S}$, we argue that it's more beneficial to model user fatigue in the frequency domain than the time domain for two reasons:
1) $\bm{S}$ only contains magnitude information of \emph{non-clicks} at each request, while in the frequency domain, we can observe both amplitude (related to magnitude) and phase (related to position) for each frequency component, which is more beneficial to capture the periodic evolution of fatigue; 
2) In the frequency domain, we can conveniently distinguish the influence of different frequency components for more elaborate modeling.

Motivated by this, we perform $N$-point FFT to transform $\bm{S}$ into the frequency domain:
\vspace{-0.3cm}
\begin{equation}
    \begin{split}
        &\bm{A}, \bm{\varphi}=FFT(\bm{S}), \\
        &\bm{A}=[A_1,...,A_k,...,A_N] \in \mathbb{R}^{N}, \\
        &\bm{\varphi}=[\varphi_1,...,\varphi_k,...,\varphi_N] \in \mathbb{R}^{N}, \\
    \end{split}
\end{equation}
where $\bm{A}$ and $\bm{\varphi}$ are the amplitude and phase vectors respectively.

To capture the different influences of different frequency components adaptively and obtain a high-order representation of amplitude and phase, we feed $\bm{A}$ and $\bm{\varphi}$ into a two-layer MLP respectively. Then, we perform an element-wise multiplication of the results to model the interaction of amplitude and phase and obtain a combination representation:
\begin{equation}
    \begin{split}
        \bm{A'}&=MLP_A(\bm{A}) \in \mathbb{R}^{N}, \\
        \bm{\varphi'}&=MLP_\varphi(\bm{\varphi}) \in \mathbb{R}^{N}, \\
        \bm{F}&=\bm{A'} \otimes \bm{\varphi'} \in \mathbb{R}^{N},
    \end{split}
\end{equation}
$\bm{F}$ contains comprehensive information about user fatigue, which is considered as a raw fatigue representation. It's noteworthy that $MLP_A$ and $MLP_{\varphi}$ are used to learn that influence of different frequency components is different. For the amplitude, components at low frequencies are more robust, while components at high frequencies may contain more noise and should be attenuated. For the phase, components at low frequencies are more important than those at high frequencies, because the position of high-frequency signals is less sensitive to phase than that of low-frequency signals.

\subsubsection{Category Modulation Net.}
Intuitively, both the upper bound of fatigue and users' patience is different for different categories. Taking inspiration from the spirit of APG \cite{yan2022apg}, we propose to generate model parameters dynamically based on different instances, which helps to capture custom patterns and enhance the model capacity. Thus, we can model user fatigue adaptively for different categories. Specifically, for each category, we formalize the upper bound as a bias vector $\bm{b_c}$, and the users' patience as a weight vector $\bm{w_c}$, which are generated from category information of the target item respectively:
\begin{equation}
    \begin{split}
        \bm{b_c}&=MLP_b(\bm{e_c}) \in \mathbb{R}^{N}, \\
        \bm{w_c}&=MLP_\alpha(\bm{e_c}) \in \mathbb{R}^{N}, 
    \end{split}
\end{equation}
where $\bm{e_c}$ is the embedding of category features, including category ID and the corresponding statistical features which can be obtained from item features $\bm{x}^i$. To model our intuition, we modulate the raw fatigue representation $\bm{F}$ as:
\begin{equation} \label{eq:cate_modulation}
    \bm{F_c}=\bm{b_c} - \bm{w_c} \otimes \bm{F} \in \mathbb{R}^{N},
\end{equation}
Compared with $\bm{F}$, $\bm{F_c}$ considers the discrepancy of fatigue across different categories and therefore represents fatigue more accurately.

\subsubsection{User Confidence Net.}
Naturally, users with high activeness generate more feedback, which makes modeling user fatigue more confident, and vice versa. Therefore, we employ a gating network to model the confidence of user fatigue according to user activeness. Specifically, we pass $\bm{e^{u}_a}$, the embedded user activeness generated from user features $\bm{x}^u$, to a MLP to produce the confidence factor $\bm{\alpha_u}$, after which element-wise multiplication is performed between $\bm{\alpha_u}$ and $\bm{F_c}$, $i.e.$,
\vspace{-0.3cm}
\begin{equation}
    \begin{split}
        \bm{\alpha_u}&=MLP_u(\bm{e^{u}_a}) \in \mathbb{R}^{N},  \\
        \bm{F_{uc}}&=\bm{\alpha_u} \otimes \bm{F_c} \in \mathbb{R}^{N}, 
    \end{split}
\end{equation}
Through the aforementioned operations, we obtain a fine-tuned fatigue representation $\bm{F_{uc}}$ which considers the category and the user biases simultaneously.

\subsection{Training}
On the top of the UIM and FRM, the user-item representation $\bm{r_{ui}}$ and the fatigue representation $\bm{F_{uc}}$ are combined by concatenation and passed to a MLP to make the final CTR prediction:
\begin{equation}
    \hat{y}=f_{\theta}(\bm{x})=MLP_o(concat(\bm{r_{ui}}, \bm{F_{uc}})),
\end{equation}
The last layer of $MLP_o$ uses \emph{Sigmoid} as activation function to project the prediction to the click probability. We adopt the widely-used logloss as the main loss, which is calculated as follows:
\begin{equation}
    L_m = - \frac{1}{|\mathcal{D}|} \sum_{(\bm{x},y) \in \mathcal{D}} (y\,log\,\hat{y} + (1-y)\,log(1-\hat{y})),
\end{equation}
where $\mathcal{D}$ denotes training set and $|\mathcal{D}|$ denotes the number of samples in $\mathcal{D}$.

Moreover, to guide the learning of FRM, we design an additional auxiliary task, $i.e.$, predicting the degree of fatigue:
\begin{equation}
    \hat{y}_f=MLP_f(\bm{F_{uc}}),
\end{equation}
where $MLP_f(\cdot)$ is a 3-layer MLP of which the last layer uses \emph{Sigmoid} as activation function. To find a confident fatigue label for $\hat{y}_f$, we take a user's behaviors in the next three days into consideration. If a user clicks on the target category, we assume that the user has not been over-exposed and mark the fatigue label as negative. If the user doesn't click the target category after a certain number of exposures, we mark the fatigue label as positive. Training samples in other situations are not used in the fatigue prediction task. By aggregating user behaviors over an appropriate time window, we can obtain a relatively stable and confident fatigue label $y_f$ on the target category. Then we calculate logloss between $\hat{y}_f$ and $y_f$, and formulate the final loss as follows:
\begin{equation}
\begin{split}
    L_f &= - \frac{1}{|\mathcal{D}|} \sum_{(\bm{x},y_f) \in \mathcal{D}} (y_f\,log\,\hat{y}_f + (1-y_f)\,log(1-\hat{y}_f)), \\
    L &= L_m + \beta L_f.
\end{split}
\end{equation}
where $\beta$ is a scaling hyperparameter that gradually increases during the training process, and the optimal maximum value of $\beta$ is determined by experiments. With the auxiliary loss $L_f$, the FRM is guided to pay attention to what users are not interested in, therefore learning user fatigue better.

\section{Experiments}

\subsection{Experimental Setup} \label{subsec:exp_setup}
\subsubsection{Datasets.}
We collect and sample online service logs\footnote{The data collection is under the application's user service agreement and users' private information is protected.} from the recommendation scenarios in Tmall Mobile between 2022/08/24 and 2022/09/26 as our experimental datasets. Then we split the data into two non-overlapped parts. The data between 2022/08/24 and 2022/09/25 is used for training while the data in 2022/09/26 is collected for testing. Table~\ref{tab:dataset} summarizes the detailed statistics of our datasets.
\begin{table}[htbp] 
\vspace{-1.0em}
\centering
    \caption{Statistics of the established dataset.}
    \vspace{-0.5em}
    \setlength{\tabcolsep}{1.5mm}{
        \begin{tabular}{c c c c c c}
            \hline
            \#Dataset & \#Users & \#Items & \#Exposures & \#Clicks & \#Purchases \\
            \hline
            train & 9.40M & 7.39M & 835.76M & 66.62M & 277.22K \\
            test & 842.71K & 2.70M & 26.62M & 2.15M & 10.82K \\
            \hline
        \end{tabular}
    }
    \label{tab:dataset}
    \vspace{-1.5em}
\end{table}

\subsubsection{Evaluation Metrics.}
\label{subsec:exps}
For the offline comparison, we use Area Under ROC Curve (AUC) as the evaluation metric. As for the online A/B testing, we use $PCTR=p(click|impression)$ and the average number of user clicks (IPV), which are widely adopted in industrial recommender systems. Moreover, we use Leaf Categories Exposed Number per user (LCEN) and Leaf Categories Clicked Number per user (LCCN) to measure the diversity of recommendations.

\vspace{-0.5cm}
\subsubsection{Competitors.}
As a representative in CTR prediction, \textbf{DIN} \cite{zhou2018deep} is chosen to be the base model. Besides, we compare the performance of our proposed FAN model with a series of state-of-the-art methods that model both \emph{clicks} and \emph{non-clicks}, $i.e.$, \textbf{DFN} \cite{xie2021deep}, \textbf{DUMN} \cite{bian2021denoising} and \textbf{Gama} \cite{xu2022gating}. Additionally, to demonstrate the effectiveness of our designed structure in FAN, we also conduct several ablation experiments:
\begin{itemize}
 \item \textbf{FAN\_w/o\_FRM (UIM)}: As a substructure of FAN, UIM can be used as a deep CTR model by adding prediction layers, which adopts the attention mechanism \cite{vaswani2017attention} to model user positive behavior sequence.
 \item \textbf{FAN\_w/o\_TFTN}: To prove the necessity of the TFTN, we remove the TFTN module and directly make use of the input fatigue time series instead.
 \item \textbf{FAN\_w/o\_CMN}: In order to verify the gain of the Category Modulation Net(CMN) of FRM to the FAN, we remove the CMN and directly process the output of TFTN via element-wise product with the output of UCN.
 \item \textbf{FAN\_w/o\_UCN}: Similar to FAN\_w/o\_CMN, we test the performance of UCN by removing it from FRM.
\end{itemize}

\vspace{-0.5cm}
\subsubsection{Implementation Details.}
All models share the same features, except that DIN ignores \emph{non-clicks} and DFN and DUMN additionally adopt explicit negative feedback of users. All the models are implemented in distributed Tensorflow 1.4 and trained with 10 parameter servers and 4 Nvidia Tesla V100 16GB GPUs. Item ID has an embedding size of 64, category ID and brand ID have an embedding size of 32 while 8 for the other categorical features. We use 8-head attention structures in UIM with a hidden size of 128 and 32-point FFT in FRM. Adagrad optimizer with a learning rate of 0.01 and a mini-batch size of 1024 is used for training. During training, $\beta$ increases linearly from 0.01 to 0.5 with training steps increasing. We report the results of each method under its empirically optimal hyper-parameters settings.

\vspace{-0.3cm}
\subsection{Overall Results}
\begin{table*}[tp] 
\centering
    \caption{Offline and online results of comparison experiments}
    \vspace{-0.5em}
    \resizebox{\textwidth}{!}{
        \begin{tabular}{l c c c c c c}
        \hline
        \multirow{2}{*}{Model} & Offline & \multicolumn{4}{c}{Online Gain} \\
        \cmidrule(r){2-2} \cmidrule(r){3-6}
        & AUC (mean$\pm$std.) & PCTR & IPV & LCEN & LCCN \\
        \hline
        DIN(Base Model) & 0.7178$\pm$0.00247 & 0.00\% & 0.00\%& 0.00\% & 0.00\% \\
        DFN & 0.7194$\pm$0.00585 & -0.30\% & -0.31\%& +0.85\% & +0.53\% \\
        DUMN & 0.7218$\pm$0.00429 & +0.17\% & -0.74\%& +0.59\% & +1.15\% \\
        Gama & 0.7225$\pm$0.00472 & +0.89\% & +0.42\%& -0.26\% & +0.57\% \\
        FAN\_w/o\_FRM(UIM) & 0.7193$\pm$0.00238 & -0.85\% & +0.79\%& +2.52\% & +1.41\% \\
        \textbf{FAN(ours)} & \textbf{0.7249$\pm$0.00176} & \textbf{+1.63\%} & \textbf{+1.09\%} &\textbf{+11.13\%} &\textbf{+3.29\%}\\
        \hline
        \end{tabular}
    }
    \label{tab:comparison}
    \vspace{-1.5em}
\end{table*}
For offline evaluation, each model has repeated five times and the best version of each model is selected for online A/B tests, which lasted 3 days from 2022/10/05 to 2022/10/08. The offline and online comparison results are presented in Table~\ref{tab:comparison} and the major observations can be summarized as follows:
\begin{enumerate}
 \item  Compared with DIN, the UIM model performs better in the offline evaluation. Although the PCTR of UIM is slightly worse, it can recommend more abundant categories than DIN and attract more IPV, implying that our design for modeling positive user interests is effective.

 \item The DFN and DUMN perform better than DIN on the offline AUC metric by distilling implicit negative feedback via explicit negative feedback. However, users' explicit negative feedback are extremely scarce in e-commerce, which is insufficient for learning negative user interests comprehensively. As a result, DFN and DUMN can't perform well in online recommendation scenarios.
 
 \item The Gama model slightly outperforms DIN, DFN, and DUMN on AUC, PCTR, and IPV by denoising implicit negative feedback in the frequency domain. However, it still focuses on modeling what users are interested in, so it's incapable of improving the diversity of recommendations. In such situations, users may be overexposed to too many similar recommendations.
 
 \item For both offline and online, our FAN model yields the best performance. It's noteworthy that FAN not only improves the efficiency of online traffic ($i.e.$, PCTR, and IPV) but also achieves impressive gains in diversity. Benefiting from the frequency-domain representation and category modulation, our modeling of fatigue is much more comprehensive and accurate, which helps users avoid seeing too many similar recommendations and explore more new items. Besides, with an elaborated auxiliary task, FAN pays more attention to what users are not interested in than the Gama model.
\end{enumerate}

\vspace{-0.3cm}
\subsection{Ablation Study}
To demonstrate the effectiveness of the designed structure in FAN, we also conduct a series of ablation experiments as detailed in Section~\ref{subsec:exp_setup}. The results are detailed in Table \ref{tab:ablation}. Totally speaking, removing FRM or key substructures of FRM from FAN leads to the decline of AUC, illustrating the effectiveness of our model design. The other supplementary conclusions are summarized as follows:
 \begin{enumerate}
  \item The comparison between FAN\_w/o\_TFTN and FAN shows that TFTN can achieve better use of the fatigue time series. With the frequency spectrum extracted by FFT, TFTN further models different impacts of different frequency components, helping to learn a better fatigue representation for the final prediction.
  \item With CMN removed, FAN\_w/o\_CMN suffers a significant performance degradation that is almost comparable to removing the whole FRM. This observation proves the correctness of our intuition that both the upper bounds of fatigue and users’ patience are different for different categories, and confirms the necessity of category modulation while modeling user fatigue.
  \item Comparing FAN\_w/o\_UCN with FAN, a degradation of AUC is observed, implying that considering user confidence is beneficial to the learning of fatigue representation.
 \end{enumerate}
\vspace{-0.1cm}
\begin{table}[!htbp] 
\vspace{-2.0em}
\centering 
    \caption{Offline results of ablation experiments}
    \vspace{-0.5em}
    \setlength{\tabcolsep}{6.0mm}{
        \begin{tabular}{l c c}
        \hline
        Model & AUC (mean$\pm$std.) \\
        \hline
        FAN\_w/o\_FRM(UIM) & 0.7193$\pm$0.00238 \\
        FAN\_w/o\_TFTN & 0.7225$\pm$0.00135 \\
        FAN\_w/o\_CMN & 0.7199$\pm$0.00147 \\
        FAN\_w/o\_UCN & 0.7237$\pm$0.00187 \\
        \textbf{FAN} & \textbf{0.7249$\pm$0.00176}\\
        \hline
        \end{tabular}
    }
    \label{tab:ablation}
    \vspace{-2.5em}
\end{table}
\vspace{-0.2cm}

\subsection{Effectiveness Analysis}
\begin{figure*}[htbp]
\vspace{-2.0em}
\begin{center}
\subfigure[TFTN output]
{
	\begin{minipage}[b]{0.44\textwidth}
	\centering
	\includegraphics[scale=0.38]{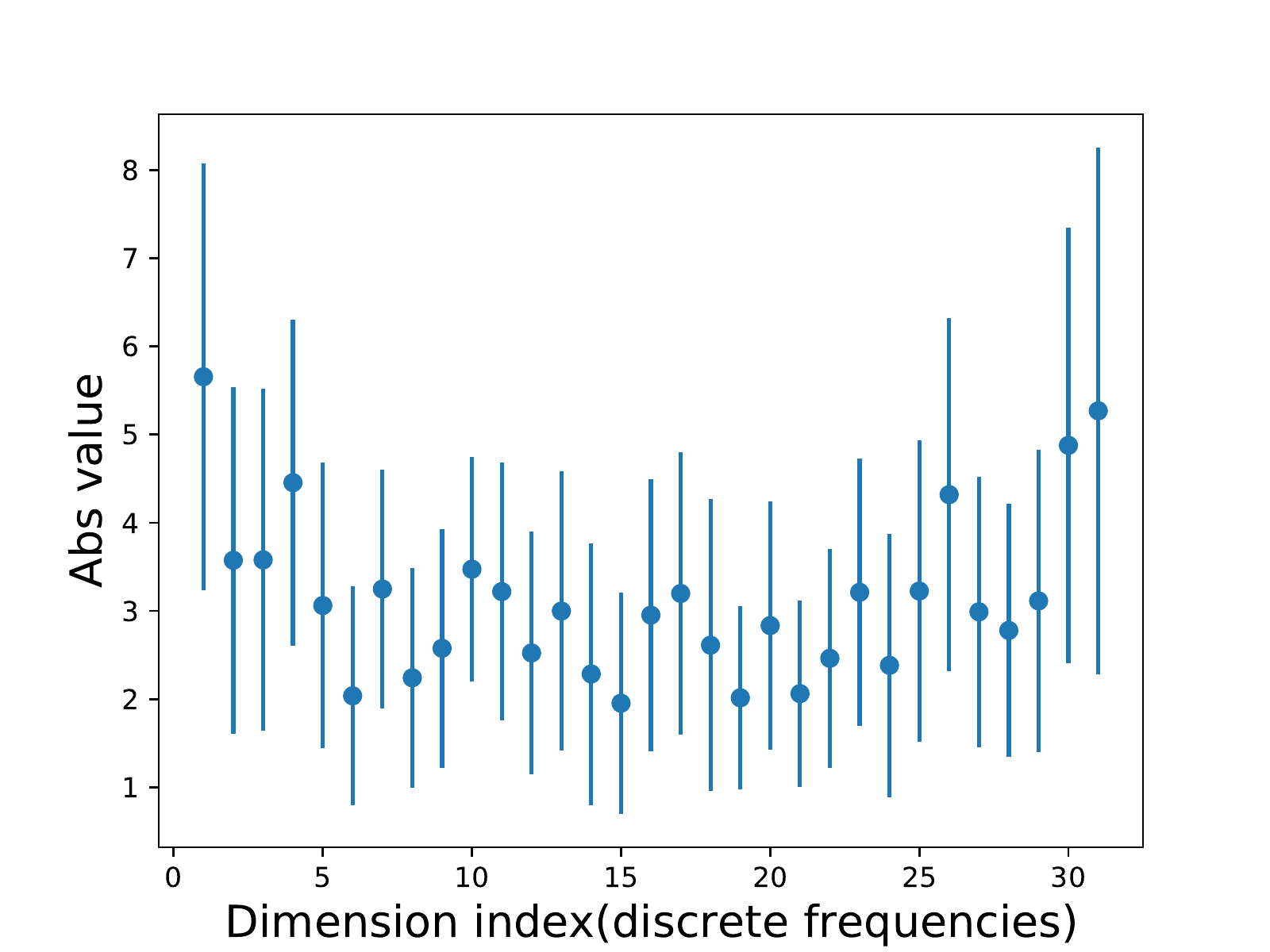}
	\end{minipage}
	\label{fig:fft_output}
}
\subfigure[FRM output]
{
	\begin{minipage}[b]{0.44\textwidth}
	\centering
	\includegraphics[scale=0.38]{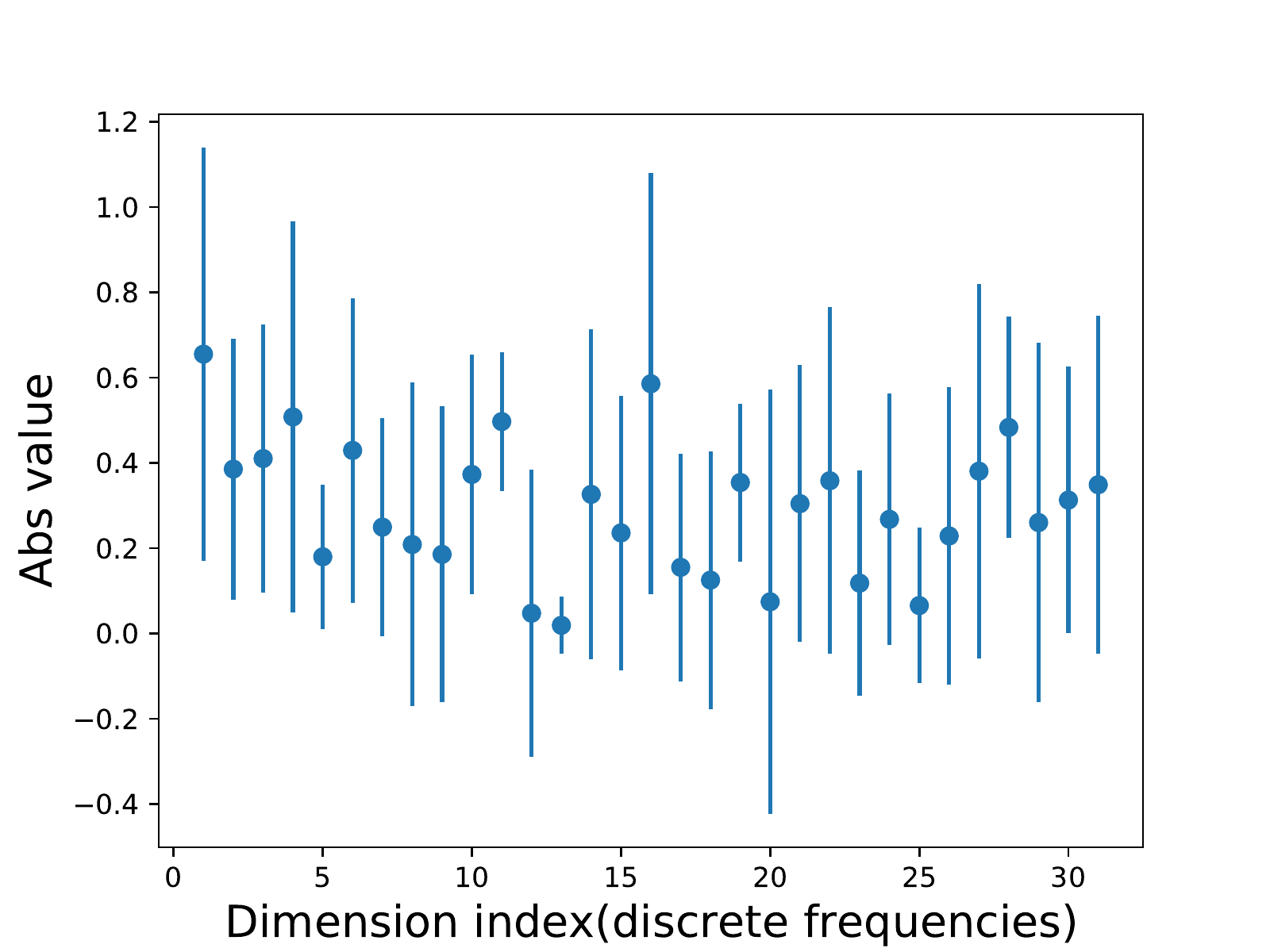}
	\end{minipage}
	\label{fig:FRM_output}
}
\end{center}
\vspace{-2.5em}
\caption{Mean and standard deviation of TFTN and FRM output}
\label{fig:Mean_std}
\vspace{-2.0em}
\end{figure*}

To further analyze how TFTN and FRM work, we visualize the output of the TFTN and FRM. Specifically, we randomly select 1000 samples and feed them to FAN, extracting $\bm{F}$ and $\bm{F_{uc}}$ for each sample. Then, we draw mean and standard deviation diagrams for $\bm{F}$ and $\bm{F_{uc}}$ respectively, as shown in Figure~\ref{fig:Mean_std}. Since the symmetry characteristic of FFT, Figure~\ref{fig:fft_output} shows an approximately symmetrical structure. Output values between different dimensions are quite different. Frequency components around the $6$-th and $15$-th dimensions have weaker responses while the other has a higher response, implying that our TFTN act as a second-order bandstop filter in FRM. From Figure~\ref{fig:FRM_output}, we can observe that the symmetrical structure no longer exists, and different frequency components are further activated. This observation implies that our devised CMN and UCN achieve the ability of frequency selection. Moreover, the standard deviation of some frequency components becomes larger while some become smaller, indicating that FRM further weakens the unimportant components and strengthens important components.

\section{Conclusion}
In this paper, we investigate the difficulties of modeling implicit negative user behaviors ($i.e.$, \emph{non-clicks}) in e-commerce and propose a novel model named FAN which captures the negative user interests in the target category from a perspective of fatigue modeling. In FAN, we apply FFT to obtain time-frequency information from the corresponding sequence of \emph{non-clicks}, which is then used to learn the high-order fatigue representation with category bias and user confidence considered. Moreover, an auxiliary task is elaborately designed to guide the learning of the FRM. In this way, a high-quality fatigue representation can be learned to facilitate improving both the CTR performance and user experience, which is validated by real-world offline datasets as well as online A/B testing. The extensive analysis further confirms the effectiveness of our model design.

%
%
%

\end{document}